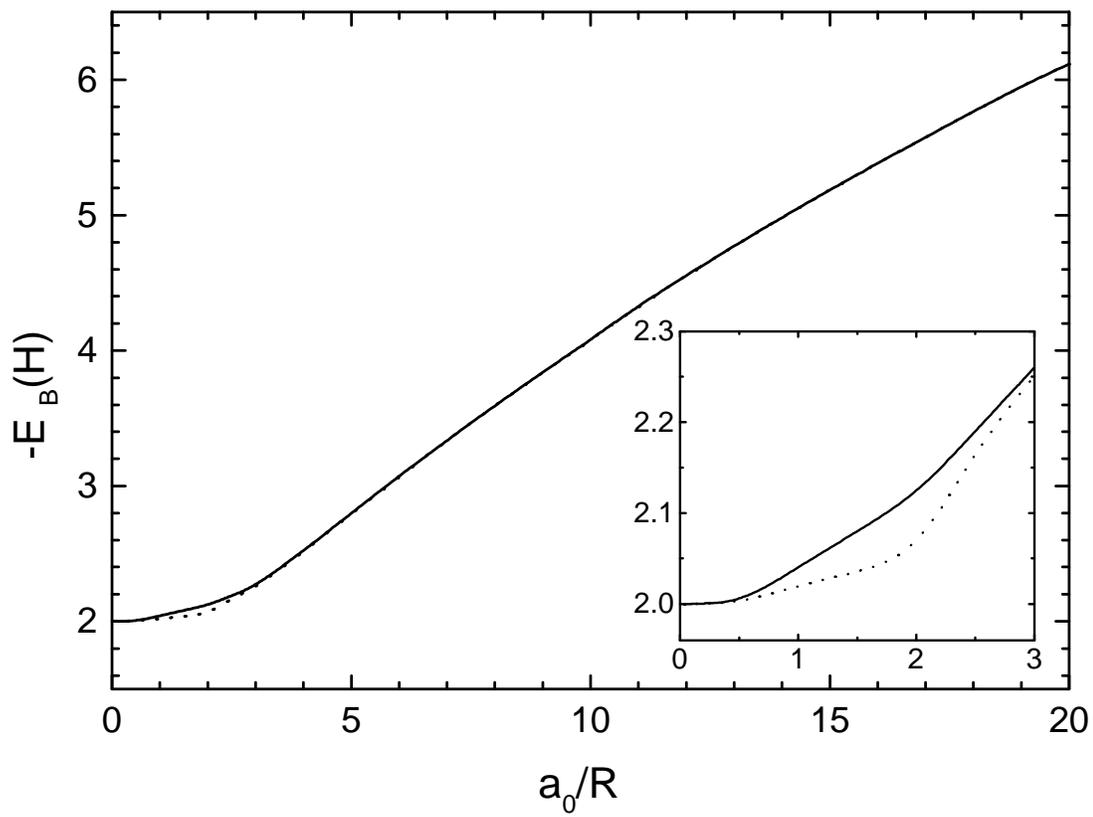

**Fig.1**

# Variational approach to the Coulomb problem on a cylinder

M. K. Kostov, M. W. Cole, and G. D. Mahan

*Department of Physics, Penn State University, University Park, Pennsylvania 16802*

## Abstract

We evaluate, by means of variational calculations, the bound state energy $E_B$ of a pair of charges located on the surface of a cylinder, interacting via a Coulomb potential $-e^2/r$. The trial wave function involves three variational parameters. $E_B$ is obtained as a function of the reduced curvature $C = a_0/R$, where $a_0$ is the Bohr radius and $R$ is the radius of the cylinder. We find that the energetics of binding exhibits a monotonic trend as a function of $C$; the known 1D and 2D limits of $E_B$ are reproduced accurately by our calculation. $E_B$ is relatively insensitive to curvature for small $C$. Its value is ~1% higher at $C=1$ than at $C=0$. This weak dependence is confirmed by a perturbation theory calculation. The high curvature regime approximates the 1D Coulomb model; within our variational approach, $E_B$ has a logarithmic divergence as $R$ approaches zero. The proposed variational method is applied to the case of donors in single-wall carbon nanotubes (SWCNTs).

## I. Introduction

The interactions of the subatomic world have been modeled by local potentials ever since the introduction of quantum physics. Some of these potentials used in quantum mechanics (QM) are exactly solvable. That is, the energy spectrum, the bound-state wavefunctions and the scattering matrix can be obtained in closed analytical form. The simplest exactly solvable potentials belong to the so-called shape-invariant class [1,2]. Most well-known potentials, like the Coulomb, Morse, Pöschl-Teller, etc. potentials have the property of shape-invariance. The Coulomb interaction in QM has been extensively studied ever since the dawn of the quantum theory, since it is the basic interaction upon which our real world is built.

An important feature of the QM models based on the Coulomb potential is the dimensionality dependence of the Coulomb interaction. The three dimensional (3D) Coulomb-Schrödinger problem is exactly solvable and it was one of the brightest examples of the success of the new quantum theory in the beginning of the twentieth century. There are many textbook applications of the 3D Coulomb model, such as the discrete energy spectrum hydrogen atom, hydrogen-like atoms, etc. Due to the spherical symmetry this problem can be reduced to one-dimensional (1D) radial problem, which can be solved exactly in terms of Laguerre polynomials [3]. The two dimensional (2D) Coulomb problem has been extensively studied in exploring the statistical mechanics of the 2D Coulomb gas [4,5]. Usually the 2D QM model consists of a (+/−) pair interacting via the 2D Coulomb potential [5] $q^2 \ln r$. In contrast to the 3D case, and in analogy with many other 2D problems, solving the 2D Schrödinger equation with that potential necessitates the use of numerical techniques. The energy spectrum is shown to obey purely discrete and semibounded in its lower part, while the wavefunctions behave like those of a simple harmonic oscillator. The 1D Coulomb-Schrödinger problem has physical relevance in the description of the linear stark effect of a 1D system [6,7]. This problem has been subject of intensive studies in the past decades and there is still some controversy in the interpretation of the results [8]. The unusual features attributed to the 1D Coulomb problem, which arise due to the $r^{-1}$–like singularity, include an infinitely bound ground state [9], degenerate eigenvalues and continuous bound-state spectrum [10]. It is believed that the usual techniques of QM alone are not sufficient for dealing with the 1D Coulomb potential. Recently, it has been shown that the complications arising due to the singularity of the 1D Coulomb problem can be avoided with the use of a generalized Coulomb potential [2].

Nowadays, there exists a wide variety of quite regular porous media having characteristic widths of the order of a nanometer. These include various carbon nanotubes materials, which can be considered to high accuracy as perfect nano-scale cylinders. Very recently, there has been a theoretical study of bound states in curved quantum layers by Duclos *et al.* [11]. They set a list of sufficient conditions to guarantee the existence of curvature-induced bound states and showed that the curvature has an essential effect on the discrete spectrum of a nonrelativistic particle constrained to a curved layer. Thus, one can address the very interesting, from practical and theoretical points of view, problem of modeling the Coulomb interactions of (+/−) pair on a nano-cylinder with a radius $R$. This problem interpolates between the 2D and 1D Coulomb problems, but does not have an analytical solution. One can anticipate a strong curvature dependence of the corresponding bound ground state. The 2D and 1D Coulomb problems can be considered as limiting cases of the cylindrical one with $R \to \infty$ and $R \to 0$, respectively. The aim of the present work is to determine, by means of variational calculations, the bound ground state energy of such a pair of charges, located on the surface of a cylinder, interacting via a Coulomb potential $-e^2/r$. We can make a reasonable guess for the trial wavefunction taking into account the exact solution for the ground state wavefunction in the 2D Coulomb problem $\Psi_{2D} = \exp(-2r/a_0)$, where $a_0$ is the Bohr radius. We choose two forms for the trial wavefunction – the first involves two variational parameters, while the second employs three variational parameters. Adding a third variational parameter to our first choice for the trial wavefunction improves the variational energy only for a narrow range of curvature. The minimum energy expectation values are obtained as a function of a curvature parameter, defined as $C = a_0/R$. The limiting case of small curvature is investigated by an alternative approach, based on a perturbation theory involving the parameter $a_0/R$. In the high curvature limit an analytical solution is obtained by using a strongly localized trial wavefunction of Gaussian type. In that case the variational approach yields ground state energy with a logarithmic singularity $\sim |\ln(a_0/R)|$.

## II. Analysis

We consider a pair of (+/−) charges, constrained to lie on the surface of a cylinder, interacting with potential of the Coulomb form. The essential results we obtain a requite general and depend only on the curvature parameter $a_0/R$. In what follows we measure the energy in modified Hartree units (1 Hartree $\equiv \mu e^4/\hbar^2$) and the lengths in modified Bohr radius units ($a_0 = \hbar^2/\mu e^2$), where $\mu$ is the reduced mass of the two-body system. The ground state energy is calculated by minimizing the expectation value of the reduced Hamiltonian $H^*$:

$$H^* = -\frac{1}{2}\left(\frac{\partial^2}{\partial z^{*2}} + \frac{1}{R^{*2}}\frac{\partial^2}{\partial \varphi^2}\right) + V(r^*),$$

$$r^* = \sqrt{z^{*2} + 2R^{*2}(1-\cos\varphi)},$$

(1)

where $\varphi$ and $z$ are the usual cylindrical coordinates, $V(r^*) \equiv -1/r^*$, $z^* = z/a_0$, $R^* = R/a_0$, $a_0 = 0.529$Å if $\mu = m_e$, which we assume to be the case here. The more general case can be obtained by appropriate scaling of our solution. In the limit of $R \to \infty$ we have an exact solution for the ground state eigenfunction in the form $\Psi_{2D} = \exp(-2r^*)$ with eigenvalue 2. The initial trial wavefunction for the cylindrical problem is chosen to be:

$$\Psi(z^*,\varphi) = \exp\left[-\alpha\left(z^{*2} + \frac{2R^2}{a_0^2}(1-\cos\varphi)\right)^{\frac{n}{2}}\right],$$

(2)

where $\alpha$ and $n$ are our variational parameters. The two-body eigenfunction $\Psi(z^*,\varphi)$ is determined by the minimization of the energy expectation value, formally written as:

$$\frac{\partial}{\partial \alpha}\left[\frac{\langle\Psi|H^*|\Psi\rangle}{\langle\Psi|\Psi\rangle}\right] = 0, \qquad \frac{\partial}{\partial n}\left[\frac{\langle\Psi|H^*|\Psi\rangle}{\langle\Psi|\Psi\rangle}\right] = 0,$$

(3)

where $\langle\Psi|H^*|\Psi\rangle$ is the expectation value of the reduced Hamiltonian. These equations are solved numerically and the ground state energy is obtained as a function of the curvature parameter $a_0/R$. We note that this form of the trial wavefunction can be improved by adding a third variational parameter $\beta$. This could be associated with either the spatial coordinate $z$ or the azimuthal angle $\varphi$. We have employed the following form for $\Psi(\alpha, \beta, n)$:

$$\Psi(\alpha,\beta,n) = \exp\left[-\alpha\left(z^{*2} + \frac{2R^2}{a_0^2}(1-\cos\varphi)\right)^{\frac{n}{2}} + \beta\cos\varphi\right]. \quad (4)$$

The ground state energy $E_B$ is shown in Fig. 1 as a function of $C$, calculated for both of the trial wave functions (Eq. 2 and Eq. 4). The energy curve has a remarkable behavior in the limit of small $C$ ($a_0/R < 1$). Surprisingly, in that limit, the energy is hardly changed by reducing $R$ from infinity to $a_0$. At $R = a_0$ the ground state energy differs from the 2D ground state value by only ~1%. On the other hand, the high curvature limit yields an infinitely bound ground state with a logarithmic singularity ~ $|ln(a_0/R)|$ as discussed below. The minimization procedure Eq. (3) determines also the optimal values of the variational parameters $\alpha$ and $n$ over the whole range of $C$. The behavior of the first parameter $\alpha$ is shown in Fig. 2. In the small $C$ limit $\alpha$ is equal to 2 (exactly the 2D result), while $\alpha$ diverges as $|ln C|$ for large $C$. Such a behavior is consistent with the numerical results for the ground state energy (Fig. 1). It indicates that the wavefunction is spread out on a scale of $a_0$ for small $a_0/R$, where the curvature has little effect on the two-body bound state, while the eigenfunction becomes increasingly localized when the curvature is large. On the other hand, the second variational parameter $n$ has different behavior from $\alpha$. For small $a_0/R$ it follows the 2D result ($n = 1$), while in the high curvature limit it increases and reaches a constant value of $1.3$, which we cannot explain. The addition of the extra variational parameter $\beta$ has little effect on the variational calculation at small and high curvature. For intermediate curvature ($a_0/R \sim 1$ to $2$) the numerical results for the ground state energy are improved by a few per cent, as indicated in Fig. 1. In that intermediate region $\beta$ has values $0.1 < \beta < 0.2$, while in the case of small and high curvature $\beta$ is very small.

Next, we test our variational calculation for small $a_0/R$. In that case a perturbation theory calculation can be carried out over the parameter $a_0/R$. We define a new variable, the arc coordinate $y^*$, as $y^* = (R/a_0)\varphi$. Expanding $\cos\varphi$ in $V(r^*)$ and assuming that $a_0/R$ is a small parameter, one can express the potential as:

$$V(r^*) = \frac{-1}{\sqrt{z^{*2} + \frac{R^2}{a_0^2}\left(1 - \frac{\varphi^2}{12} + ...\right)}} \cong \frac{-1}{\sqrt{z^{*2} + y^{*2}\left(1 - \frac{a_0^2 y^{*2}}{12R^2} + O\left(\frac{a_0^4}{R^4}\right)\right)}} \quad (5)$$

Consequently, the dimensionless Schrödinger equation becomes:

$$\left\{\frac{\partial^2}{\partial z^{*2}} + \frac{\partial^2}{\partial y^{*2}} + \frac{2}{\sqrt{z^{*2}+y^{*2}}}\left[1 + \frac{a_0^2 y^{*4}}{24R^2(z^{*2}+y^{*2})}\right] + E\right\}\Psi = 0, \qquad (6)$$

where the reduced energy is expressed in units of modified Hartrees. Further, we can relate this equation to the 2D Schrödinger equation by introducing the polar coordinates $\rho^* = \sqrt{z^{*2}+y^{*2}}$; $\theta = arctg(y^*/z^*)$. Then, in the small curvature limit, the problem on can be reduced to:

$$\left[\frac{1}{\rho^*}\frac{\partial}{\partial \rho^*}\rho^*\frac{\partial}{\partial \rho^*} + \frac{2}{\rho^*}\left(1 + \frac{a_0^2 \rho^{*2}\sin^4\theta}{24R^2}\right) + E\right]\Psi = 0, \qquad (7)$$

$$v(\rho^*) = v_{2D}(\rho^*) + \delta v(\rho^*) = -\frac{1}{\rho^*} - \frac{a_0^2 \rho^{*2}\sin^4\theta}{24R^2}. \qquad (8)$$

The unperturbed ground state eigenfunction is the 2D wavefunction $\Psi_{2D} = exp(-2\rho^*)$, while the perturbative potential is $\delta v(\rho^*)$. Performing first order perturbation theory, we obtain an analytical expression for the ground state energy in the form:

$$E_0^{cyl} = E_{2D} + \delta E^{(1)} = -2\left[1 + \frac{1}{256}\left(\frac{a_0}{R}\right)^2\right]. \qquad (9)$$

Comparison between the variational and perturbation theory calculations, in the range of small $a_0/R$, is shown in the inset of Fig.3. The first order perturbation theory, which constitutes an upper limit to the bound state energy, is in good agreement with our variational theory results. It differs from the variational approach only by ~1% and confirms the initially surprising result of a weak dependence of the energy on the curvature at small $a_0/R$.

In the remainder of this section we analyze the behavior of the ground state energy in the high curvature limit. In order to achieve an analytical approximation to the numerical results, we replace the trial wavefunction (Eq.2) with a trial function of Gaussian type:

$$\Psi_{HC}(z^*) = \exp\left[-\alpha z^{*2}\right], \qquad (10)$$

where $\alpha$ is a variational parameter. This form of the trial wavefunction does not include $\varphi$ dependence, because at large $a_0/R$ the azimuthal angle dependence in Eq.2 is very weak and we reach effectively the 1D Coulomb problem. Such a choice is reasonable, since in that range of $a_0/R$ the ground state function should be a highly localized function of $z^*$. Further, we tested $\Psi_{HC}(z^*)$ versus our first choice of trial function (Eq. 2) for $a_0/R>10$. The Gaussian wave function (Eq. 10) yielded energies which were only a few percent different from the numerical results obtained with our initial trial wave function. On the other hand, employing $\Psi_{HC}(z^*)$ we can reach a simple analytical expression for the bound energy. More precisely, in our variational calculation we first integrate over the angle $\varphi$ and obtain a complete elliptic integral of the first kind [12] $K(k)$, with $k=(z^{*2}+4R^2/a_0^2)^{-1/2}$. Next, we expand $K(k)$ in powers of $k$ and perform the integration over $z^*$, which yields a confluent hypergeometric function [12]. Keeping only the terms of zero order in $R^2/a_0^2$ we get:

$$\langle E_0 \rangle_{HC} = \frac{\alpha}{2} + \sqrt{\frac{2\alpha}{\pi}}\left[\ln\alpha - 2\ln\left(\frac{a_0}{R}\right) - 0.384\right]. \tag{11}$$

Applying the minimization procedure Eq. 3 for $\Psi_{HC}(z^*)$ we obtain the optimal values of $\alpha$ over the entire range of $a_0/R$; the results are shown in Fig. 2. Numerical fit of the optimized $\alpha$ in the case of high $a_0/R$ can be obtained:

$$\alpha \xrightarrow[\frac{a_0}{R}\to\infty]{} -17.1 + 7.2\ln\left(\frac{a_0}{R}\right) \tag{12}$$

That behavior of $\alpha$ leads to a logarithmic singularity $\sim |ln(a_0/R)|$ of the ground state energy as $a_0/R$ increases. In 1D the potential energy expectation value diverges for a $1/r$ potential and thus the spectrum is unbounded; the particle "falls to the center of the attractive force". In the cylindrical case, as $R$ approaches zero, the ground state has a logarithmic divergence and the singularity at $R=0$ can be avoided by introducing a cut off radius $R_0$. To conclude this section we compare the ground state energies computed from $\Psi_{HC}(z^*)$ and $\Psi(\alpha,\beta,n)$ on Fig.3. We find that $\Psi_{HC}(z^*)$ is a very good approximation for the trial wavefunction in case of large $a_0/R$.

## III. Donors on single-wall carbon nanotubes

One application of the present model is donors in single-wall carbon nanotubes (SWCNTs). Nitrogen is the most prevalent donor in most tubes. It is a substitutional impurity and acts as a singly charged donor. SWCNTs have a variety of energy band structures, depending upon the chirality of the tube. About one-third are metallic and the remainder are semiconducting. Earlier calculations [13-15] have determined the location of the donors states in metallic tubes using numerical methods. Here we try to determine the location of the donors using the effective mass approximation. Both metallic and semi-conducting tubes have numerous conduction band minima. The effective mass is determined by the band curvature near the band minimum. Each conduction band minimum has a donor state whose value is determined by the effective mass and dielectric constant. The effective mass is given by the curvature in the band at the minimum energy point. The choice of the dielectric function is a problem. For a single, isolated tube, the background dielectric constant should be one. However, most measurements are made in ropes or bundles of tubes, where there will be an average dielectric constant, similar in value to that found in graphite. The same considerations apply to holes bound to acceptor states. Each band extremum has a donor or acceptor bound state. These states may overlap in energy with states from other bands. In that case the state is a scattering resonance rather than a true bound state. In metallic tubes, all of the donor and acceptor states are resonances. That is the case for the resonances reported in Refs. [13,14].

In this work we consider (10,10) armchair carbon nanotubes, which have a radius of $R$=6.8 Å. The energy bands for the (10,10) tube [16], based upon the tight binding approximation, are shown in Fig. 4. A distinctive feature for all armchair tubes is the band degeneracy between the highest valence band and the lowest conduction band at $k=\pm 2\pi/(3a)$, where the bands cross the Fermi level. For a (10,10) tube there are four conduction band minima (Fig. 4), which determine the location of the donor (acceptor) bound states. To obtain the binding energy of these states we apply the variational approach developed in Section 2, using the effective mass approximation and assuming that the Coulomb interaction is scaled by the dielectric constant of graphite $\tilde{\varepsilon}=3$. We find that all donor (acceptor) bound states "fall" very close to the Fermi level (the band crossing point). Thus, we conclude that upon doping the SWCNTs with electron donors and acceptors, the conductivity of SWCNTs' ropes will increase significantly.

## IV. Summary

We have investigated the QM Coulomb problem of a pair of (+/−) charges, located on a cylinder, by means of variational calculations. The ground state is obtained as a function of a curvature parameter, defined as $a_0/R$. Surprisingly, in the limit of small curvature ($0<a_0/R<1$) the bound energy is a very weak function of $a_0/R$. This result is confirmed by a perturbation theory calculation. In the high curvature limit, an alternative variational approach, which uses a trial function of Gaussian type, can be applied to yield an analytical solution for the ground energy. The main result of the present work has been to show that the solution of the Coulomb problem on a cylinder can be divided into two distinctive regimes of behavior – small and high curvature. The Coulomb problem, in the small curvature limit, closely approximates the 2D Coulomb problem, while the high curvature limit can be associated with the 1D Coulomb model, where the ground state energy diverges as $\sim |ln(a_0/R)|$. Within our method there remain unanswered questions concerning the application of the proposed model to real systems and the excitation spectrum. These will be addressed in future work.

We would like to thank Susana Hernandez, L. W. Bruch and M. L. Classer for helpful discussions. This work was supported by the Army Research Office and the Petroleum Research Fund of the American Chemical Society. Milen Kostov is grateful to Air Products and Chemicals, Inc. (APCI) for its support through APCI/PSU graduate fellowship.

**Figure captions**

**Fig. 1** Ground state energy of a pair of charges, constrained on a cylinder, as a function of the curvature parameter $a_0/R$. The energy is in modified Hartree units, described in the text. The solid line is the variational theory result obtained from a trial wave function $\Psi(\alpha, \beta, n)$ given by Eq.4; the dotted line corresponds to the variational results obtain from a trial wave function $\Psi(z^*)\varphi$ (Eq.2), i.e., choosing $\beta=0$.

**Fig. 2** $\alpha$ variational parameter as a function of $a_0/R$. The solid line is the variational theory result obtained from trial wave function given by Eq.2; the dashed line corresponds to the variational result with a Gaussian trial wave function $\Psi_{HC}(z^*)$ (Eq.10).

**Fig. 3** Comparison of variational theory calculations with two different trial wave functions. The solid line corresponds to the wave function given from Eq.4; the "up-triangle" curve corresponds to the Gaussian trial wavefunction. On the inset graph, a comparison between variational and perturbation theory calculations is shown. The solid line is the variational theory result obtained from a trial wave function given by Eq.4; the dashed line corresponds to the perturbation theory (PT) results.

**Fig. 4** 1D energy dispersion relations for (10,10) armchair tube; $a=2.46$Å is the graphite lattice constant, $\gamma_0 \sim 3$eV is the nearest-neighbor carbon-carbon overlap integral[16].

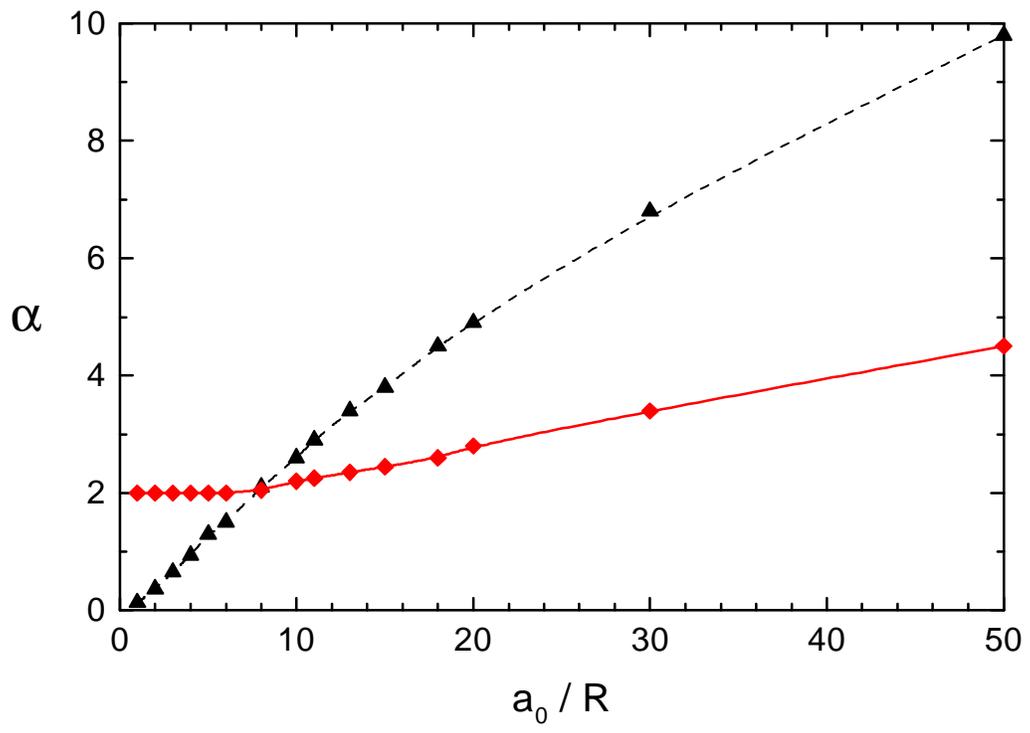

**Fig. 2**

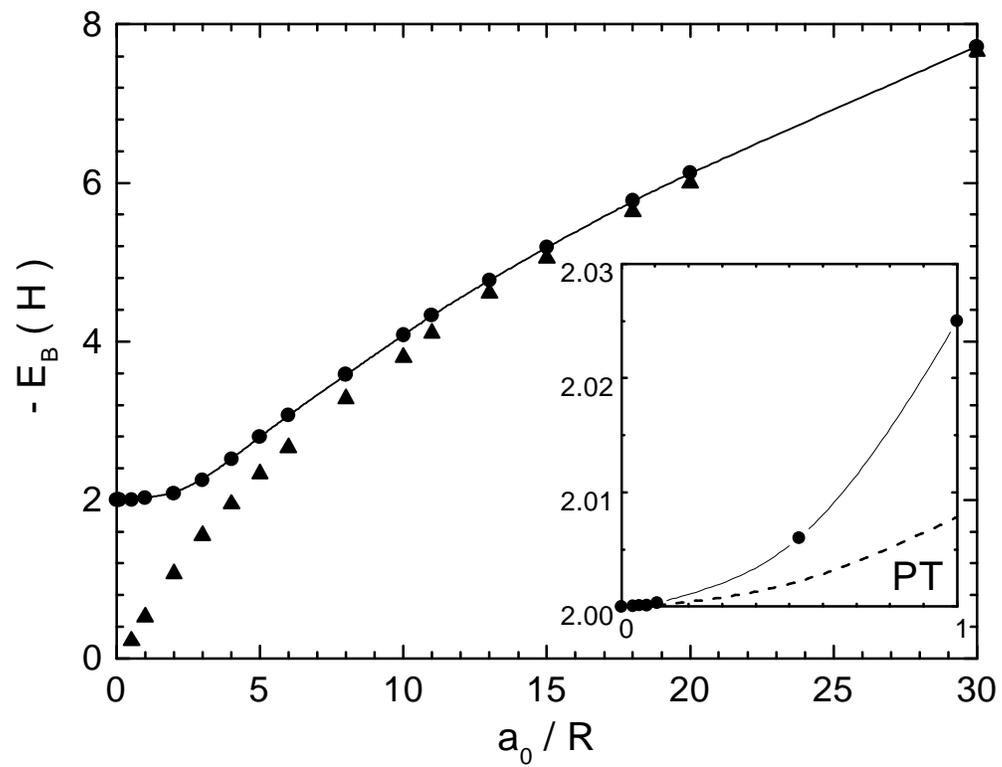

**Fig. 3**

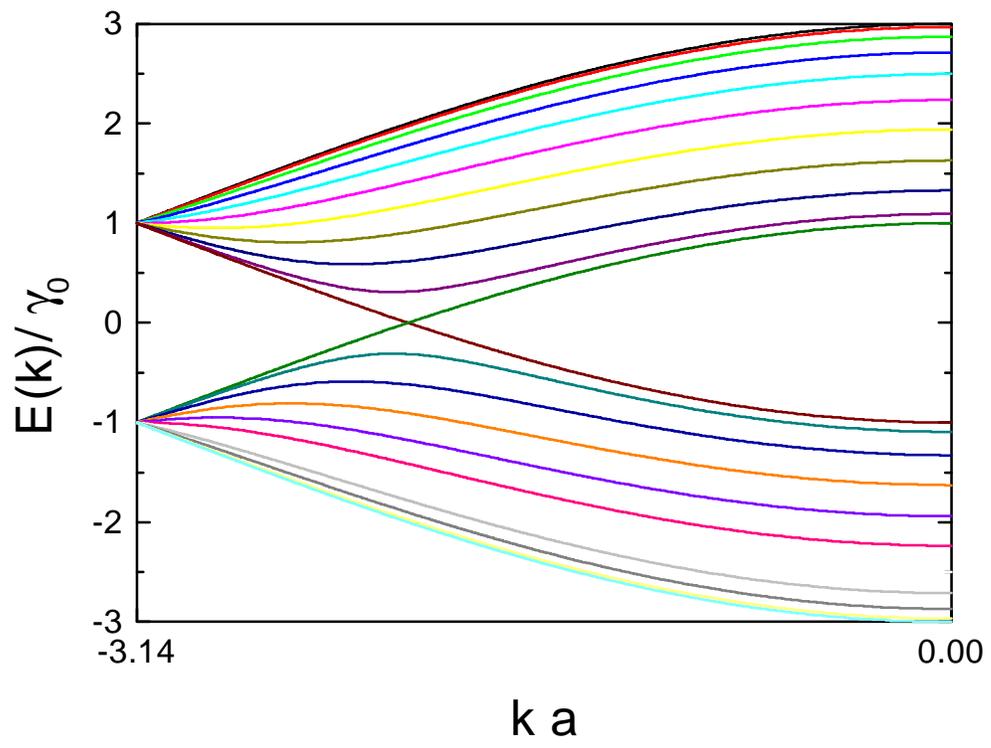

**Fig. 4**